\documentclass[twocolumn,preprintnumbers,amsmath,amssymb,bibnotes]{revtex4-1}

\usepackage{graphicx}% Include figure files
\usepackage{dcolumn}% Align table columns on decimal point
\usepackage{bm}% bold math
\usepackage{cancel} % cross out terms

\usepackage{float}
\restylefloat{figure}
\usepackage{epstopdf}

\newcommand{\be}{\begin{equation}}
\newcommand{\ee}{\end{equation}}
\newcommand{\comment}[1]{}
\newcommand{\req}[1]{(\ref{#1})}
\def\vev#1{\langle #1 \rangle}
\def\ket#1{[ #1 ]}
\def\si{\sigma}
\def\eps{\epsilon}
\def\fc#1#2{\frac{#1}{#2}}

\def\ist{\!\!=\!\!}

\newcommand{\nwc}{\newcommand}
\nwc{\ba}  {\begin{array}}
\nwc{\ea}  {\end{array}}

\nwc{\bea} {\begin{eqnarray}}
\nwc{\eea} {\end{eqnarray}}
\nwc{\nn} {\nonumber}
\nwc{\nnn} {\nonumber \vspace{.2cm} \\ }

\nwc{\bda} {\bdm\ba{lcl}}
\nwc{\eda} {\ea\edm}

\def\vev#1{\langle #1 \rangle}
\def\ket#1{[ #1 ]}

\nwc{\ds}  {\displaystyle}

\nwc{\ra}{\rightarrow}
\nwc{\lra}{\longrightarrow}
\def\lf{\left}\def\ri{\right}

\nwc{\p} {\partial}
\nwc{\Tr}{{\rm Tr}}
\def\IR{{\bf R}}

\def\ap{\alpha'}

\def\rng{\rangle}
\def\Mc{{\cal M}}
\def\Nc{{\cal N}}

\def\ov{\overline}

\def\FF#1#2{{_#1F_#2}}

\def\IC{{\bf C}}\def\IN{{\bf N}}

\begin{document}

\preprint{MPP--2009--158}

\title{On tree--level higher order gravitational couplings in superstring theory}

%%%%% \author{\ldots}
%%%%% \affiliation{\ldots}

\author{Stephan Stieberger}
\affiliation{Max--Planck--Institut f\"ur Physik, Werner--Heisenberg--Institut,
80805 M\"unchen, Germany}
% \textbackslash\textbackslash


\begin{abstract}
We consider the scattering amplitudes of five and six gravitons at tree--level 
in superstring theory.
Their power series expansions in the Regge slope $\alpha'$ are analyzed through the order $\alpha'^8$ showing some interesting constraints on higher order gravitational 
couplings in the effective superstring action like  the absence of $R^5$ terms.
Furthermore, some transcendentality constraints on  the  coefficients
of the non--vanishing couplings 
are observed: the absence of zeta values of even weight through the order $\ap^8$ like 
the absence of $\zeta(2)\zeta(3) R^6$ terms.
Our analysis is valid for any superstring background in any space--time dimension, 
which allows for a conformal field theory description.
\end{abstract}

%\pacs{11.25.Wx, 11.30.Pb, 12.38.Bx}% PACS, the Physics and Astronomy
\maketitle

Superstring theories contain a massless spin--two state identified as a graviton.
Its interactions  are studied by graviton scattering amplitudes.
Due to the extended nature of strings the latter are generically 
non-Ðtrivial functions on the string tension $\ap$. 
In the effective field theory (FT) description this $\ap$-Ðdependence
gives rise to a series of infinite many higher order gravitational couplings
governed by positive integer powers in $\ap$.
The classical Einstein--Hilbert term 
 is reproduced in the zero--slope limit $\ap\!\!\ra\!\!0$.
The modification of the Einstein equations  through the order
$\alpha'^3$  has been 
derived by studying tree--level four graviton scattering 
amplitudes \cite{Gross:1986iv,Gross:1986mw,Kikuchi:1986rk,Cai:1986sa} or
alternatively by computing four--loop $\beta$--functions of the underlying $\sigma$--model \cite{Grisaru:1986kw,Grisaru:1986dk}.
Up to this order the effective (tree--level) superstring action focusing on pure gravitational bosonic terms reads in $D$ space--time dimensions
\be\label{LAG}
{\cal L}_{\rm tree}=\fc{1}{2\kappa^2}\ R+\fc{\ap^3}{2^9\; 4!\;\kappa^2}\  \zeta(3)\  t_8t_8\ R^4\ ,
\ee
with the gravitational coupling constant $\kappa$ in $D$ dimensions, 
the Riemann scalar $R$, the Riemann tensor $R_{\mu\nu\rho\si}$  and the 
tensor $t_8$ defined in equation (4.A.21) of \cite{Schwarz}: 
% $t_8t_8R^4\equiv t_8^{\mu_1\mu_2\ldots\mu_8}t_8^{\nu_1\nu_2\ldots\nu_8}
% \prod_{i=1}^4 R_{\mu_i\mu_{i+1}\nu_i\nu_{i+1}}$.
\bea\label{Riemann}
t_8t_8R^4&\equiv& t_8^{\mu_1\mu_2\ldots\mu_8}t_8^{\nu_1\nu_2\ldots\nu_8}\\
&\times& R_{\mu_1\mu_2\nu_1\nu_2}R_{\mu_3\mu_4\nu_3\nu_4}R_{\mu_5\mu_5\nu_5\nu_6}
R_{\mu_7\mu_8\nu_7\nu_8}\ .\nn
\eea
If the indices are restricted to  $D\!=\!4$ the combination 
$t_8t_8R^4$ becomes the Bel--Robinson tensor.
The absence of $R^2$ and $R^3$ terms in superstring theory is shown in  
\cite{Metsaev:1986yb}.

In $D\ist4$ the result  simply follows by expanding  the 
(only independent and non--vanishing) four--graviton amplitude 
\be\label{fourgr}
\Mc(1^-,2^-,3^+,4^+)\ist\lf(\fc{\kappa}{2}\ri)^2\fc{\vev{12}^8
{\ket{12}}}{N(4)\vev{34}}\fc{B(s_{12},s_{14})}{B(-s_{12},-s_{14})}
\ee
through the order $\ap^3$. The superscripts $\pm$  denote the helicities of 
the corresponding gravitons. 
Above, we have introduced the kinematic 
invariants $s_{ij}\!=\!2\ap k_ik_j$ involving the external (on--shell) momenta $k_i$ 
and the Euler Beta function $B$ encoding the $\ap$--dependence 
of the full string amplitude. 
Furthermore, $\vev{ij},\ket{ij}$  are spinor products (see e.g. 
\cite{Mangano:1990by,Dixon:1996wi}), 
$s_{ij}~\ist~\ap~\{i,j\}\!:=\!\ap \vev{ij}\ket{ji}$, and  
$N(n)=\prod_{i=1}^{n-1}\prod_{j=i+1}^n\vev{ij}$~\cite{Berends:1988zp}.

In Eq. \req{LAG} further $\alpha'$--corrections ${\cal L}'_{\rm tree}$  arise from terms with  higher powers in the Riemann tensor $R_{\mu\nu\rho\si}$ supplemented by covariant derivatives $D$. In the following these terms are collectively denoted by $t_{m,n}D^m R^n$, with some tensor $t_{m,n}$ contracting $D$ and the Riemann tensor $R$.
Generically, the set of these additional interactions may be summarized in the series
\bea
{\cal L}'_{\rm tree}&=&\kappa^{-2}\sum_{n\geq4}^\infty\sum_{m=0}^\infty
\ap^{n-1+m}\hskip-0.15cm\sum_{i_r\in\IN,i_1>1 \atop i_1+\ldots+i_d=n-1+m}'
\zeta(i_1,\ldots,i_d)\nn\\
&\times&\ds{c_{m,n,\bm{i}}\ \ t^{\bm{i}}_{m,n}D^{2m} R^n\ ,}
\label{generic}
\eea
with multi--zeta values (MZVs) 
$$\zeta(i_1,\!\ldots,\!i_d)=
\sum\limits_{n_1>\ldots>n_d>0}\ \prod\limits_{r=1}^dn_r^{-i_r},\ \ \ i_r\in\IN,\ i_1>1$$
of transcendentality degree $\sum_{r=1}^di_r\ist n-1+m$ and depth $d$ 
supplemented by some rational coefficients  $c_{m,n,\bm{i}}$.
The prime at the sum \req{generic} means, that the latter runs only over  
a basis of independent MZVs of weight $n-1+m$ \cite{commhigh,Zagier,shuffle,Landon}.
For a recent account on MZVs see Ref. \cite{Blumlein:2009cf}.

The terms in the sum \req{generic} are probed by computing scattering amplitudes 
of $n$ gravitons and analyzing their power series in $\ap$ \cite{advantage}. 
For $n\ist4$ it is straightforward to extract the relevant information, 
since in this case each term in the $\ap$--expansion of \req{fourgr} 
directly translates into a term in the effective action \req{generic}. 
Moreover, the MZV coefficients of the latter are simply products of 
Riemann zeta functions $\zeta(i_1)$ of odd degree $i_1$ as a consequence of: 
\be\label{identzeta}
\fc{B(s_{12},s_{14})}{B(-s_{12},-s_{14})}\ist-e^{-2\!\sum\limits_{n=1}^\infty
\fc{\zeta(2n+1)}{2n+1}(s_{12}^{2n+1}+s_{13}^{2n+1}+s_{14}^{2n+1})}.
\ee
In this note we discuss the consequences of  (on--shell) superstring scattering of five and six gravitons to  the correction terms~\req{generic}.
We find, that some of the coefficients $c_{m,n,\bm{i}}$ are vanishing
and the MZVs of the non--vanishing terms follow some specific pattern.
%%  Tree--level scattering of four gravitons in superstring theory has been performed 
%% in \cite{Gross:1986iv,Gross:1986mw} and determines the second term of \req{LAG}.

The string world--sheet describing the tree--level string $S$--matrix of $N$ 
gravitons is described by a complex sphere with $N$ (integrated) insertions $z_i$ 
of graviton vertex operators $V_G(\eps,\ov z_i,z_i)$:
\bea\label{Start}
&&\ds{\Mc(1,\ldots,N)=V_{CKG}^{-1}}\\
&&\hskip0.25cm\ds{\times
\lf(\prod_{j=1}^N\int_{\IC}d^2z_j\ri) \vev{V_G(\eps_1,\ov z_1,z_1)\ldots 
V_G(\eps_N,\ov z_N,z_N)}.}\nn
\eea
Here, the factor $V_{CKG}$ accounts for the volume of the conformal Killing group.
One of the key properties of graviton amplitudes in string theory is that 
at tree--level they can be expressed as sum over squares of (color ordered) 
gauge amplitudes in the left-- and right--moving sectors. 
This map, known as Kawai--Lewellen--Tye (KLT) relations \cite{Kawai}, 
gives a relation between 
a closed string tree--level amplitude on the sphere and a sum of squares of (partial ordered) open string disk amplitudes.  
On the  string world--sheet  of the sphere describing the $N$--graviton amplitude
\req{Start}
the KLT relations are a consequence of  decoupling  holomorphic and anti--holomorphic
sectors by splitting the complex sphere integration over the coordinates 
$\ov z, z\!\in\!\IC$ into two real ones $\eta,\xi\!\in\!\IR$ 
describing products of two open string disk amplitudes.
At the level of degrees of freedom this is anticipated from the fact, that 
the graviton vertex operator $V_G$ splits into a product of
non--interacting open string states describing the vertex operator of 
massless vectors $V_g$, i.e.:
$V_G(\eps,\ov z,z)\simeq V_g(\ov\eps,\eta)\ \times V_g(\eps,\xi)$, 
subject to the decomposition of polarization tensors
$\eps_{\mu\nu}\ist\ov\eps_\mu\otimes\eps_\nu$. In the graviton amplitude \req{Start}
the latter comprises into the linearized Riemann tensor 
$R_{\mu\nu\rho\sigma}\ist\kappa\ k_{[\mu}k_{[\rho}
\ \ov\eps_{\nu]}\otimes\eps_{\sigma]}$.

For the two cases $N\ist5,6$, which  we shall consider 
in the sequel, we have the following relations
(with  $\Mc(1,\ldots,N)\ist(\fc{\kappa}{2})^{N-2}M(1\ldots N)$) \cite{Kawai}:
\begin{widetext}
\bea
M(12345)&=&\ds{(2\ap\pi)^{-2}\ \sin(\pi s_{12})\ \sin(\pi s_{34})\ \bar A(12345)\ 
A(21435)+(\overline{23}),}\label{klti}\\
\ds{M(123456)}&=&\ds{(2\ap\pi)^{-3}\sin(\pi s_{12})\sin(\pi s_{45})
\bar A(123456)}\lf\{\sin(\pi s_{35})A(215346)+\sin[\pi(s_{34}+s_{35})]
A(215436)\ri\}\!+\!(\overline{234}).\ \ \ \ \label{kltii}
\eea
\end{widetext}
The KLT relations (\ref{klti},\ref{kltii})
hold for any superstring background and are insensitive
to the compactification details or the amount of supersymmetries.
The gluon amplitudes $A(1\ldots N)$ are the color--ordered subamplitudes of the underlying gauge theory. 
%% They give rise to the full $N$--point gauge amplitude $\Ac$ 
%% according to the decomposition ($i_\si:=\si(i)$)
%% \be\label{colordecomp}
%% $$\Ac(1,\ldots,N)=g_{YM}^{N-2}\hskip-0.4cm\sum_{\sigma\in S_{N}/\IZ_{N}}
%% \hskip-0.3cm\Tr(T^{a_{1_\si}}\ldots T^{a_{N_\si}})\; A(1_\si\ldots N_\si),$$
%%  with $g_{YM}$  the gauge coupling constant in $D$ space--time dimensions \cite{comment}.
Hence, in superstring theory the tree--level computation of graviton amplitudes 
boils down to considering squares of tree--level gauge amplitudes $A$.
For this sector explicit computations have been performed and results are 
accessible for the cases $N\ist4$ \cite{Schwarz,GS}, 
for $N\ist5$ \cite{Medina:2002nk,DAN,STi,STii},
for $N\ist6$ \cite{DAN,STi,STii,Potsdam} and $N\ist7$ \cite{WARD}.
Moreover, through the order $\ap^2$ the full $N$--gluon MHV--amplitude
is given in \cite{STi,STii}, while the order $\ap^3$ has been constructed in 
\cite{Boels:2008fc}.
Based on all these results and the relations (\ref{klti},\ref{kltii}) the five-- and six--graviton amplitudes can be derived.

The result for the five--gluon subamplitude  can be given for 
any space--time dimension $D$ in the form \cite{Barreiro:2005hv}
\be\label{Brazil}
A(12345)=C_1\ A_{YM}(12345)+C_2\ A_{F^4}(12345)\ ,
\ee
with the kinematical factors encoding the pure YM part $A_{YM}(12345)$ and 
the genuine string part $A_{F^4}(12345)$
\begin{widetext}
\bea\label{kini}
A_{F^4}(12345)&=&(2\ap)^2\ \{\ K(\eps_1,\eps_2,\eps_3,k_3,\eps_4,k_4,\eps_5,k_5)
+(k_1k_2)^{-1}\ \lf[\ (\eps_1\eps_2)\ 
K(k_1,k_2,\eps_3,k_3,\eps_4,k_4,\eps_5,k_5)\ri.\nn\\
&&+\lf.(\eps_1k_2)\ K(\eps_2,k_1+k_2,\eps_3,k_3,\eps_4,k_4,\eps_5,k_5)-(\eps_2k_1) 
\ K(\eps_1,k_1+k_2,\eps_3,k_3,\eps_4,k_4,\eps_5,k_5)\ \ri]\nn\\
&&+\ {\rm cyclic\ permutations}\ \}\ ,\ {\rm with:}\ \  
K(\eps_1,k_1,\eps_2,k_2,\eps_3,k_3,\eps_4,k_4)= t_8^{\mu_1\ldots\mu_8}
\eps_1^{\mu_1}k_1^{\mu_2}\hskip-0.15cm\ldots\eps_4^{\mu_7}k_4^{\mu_8}.
\eea
\end{widetext}
Furthermore, we have the two (basis) functions  
\be
C_1=s_2s_5\ f_1+(s_2s_3+s_4s_5)\ f_2\ \ \ ,\ \ \ C_2=f_2\ ,
\ee
defined by the two hypergeometric functions~$\FF{3}{2}$:
\bea
&\hskip-0.3cm\ds{f_1=\hskip-0.1cm\int_0^1\hskip-0.3cm dx\hskip-0.15cm 
\int_0^1\hskip-0.3cm dy\ x^{s_2-1}y^{s_5-1}(1-x)^{s_3}(1-y)^{s_4}(1-xy)^{s_{35}}\ ,}
\nonumber\\
&\hskip-0.8cm\ds{f_2=\hskip-0.1cm\int_0^1\hskip-0.3cm dx\hskip-0.15cm
\int_0^1\hskip-0.3cm dy\ x^{s_2}y^{s_5}(1-x)^{s_3}(1-y)^{s_4}(1-xy)^{s_{35}-1}\ .} 
\label{FUNC}
\eea
In $D\ist4$ spinor notation the expressions in \req{Brazil} 
boil down to the maximally helicity violating YM five--point  partial 
amplitude~\cite{Parke:1986gb,Berends:1987me}
\be\label{pt}
\hskip-0.15cm A_{YM}(1^-2^-3^+4^+5^+)=i\ 
\frac{\langle 12\rangle^4}{\langle 12\rangle\langle 23\rangle \langle 34\rangle
\langle 45\rangle \langle 51\rangle}\ ,
\ee
and to the term
\bea\label{also}
A_{F^4}(1^-2^-3^+4^+5^+)&=&\ap^2\;
\lf(\ \vev{12}\ket{23}\vev{34}\ket{41}\ri.\\
&-&\lf.\{3,4\}\{4,5\}-\{1,2\}\{1,5\}\ \ri)\nn\\
&\times&A_{YM}(1^-2^-3^+4^+5^+)\nn
\eea
describing the leading string correction to the YM amplitude~\req{pt}. 
In the  effective action \req{also}  gives rise to 
the contact term~$t_8 F^4\!\equiv\! t_8^{\mu_1\nu_1\ldots\mu_4\nu_4}
F_{\mu_1\nu_1}F_{\mu_2\nu_2}F_{\mu_3\nu_3} F_{\mu_4\nu_4}$ with the field strength 
$F_{\mu\nu}$. 

Using in \req{klti} the expression \req{Brazil} yields the full five--graviton amplitude.
Its $\ap$--expansion is determined by  expanding
the basis \req{FUNC}, which is achieved with \cite{Huber:2005yg}.  
In the $D\ist4$ FT--limit, with 
$f_1\!\!\stackrel{\ap\ra0}{\ra}\!\fc{1}{s_2s_5}$ and $f_2\!\!\stackrel{\ap\ra0}{\ra}\!0$, the only independent five graviton
amplitude reduces to~\cite{Berends:1988zp}

\vskip-0.5cm
\be\label{find50}
\lf.M(1^-2^-3^+4^+5^+)\ri|_{\ap^0}=-4i\ \fc{\vev{12}^8}{N(5)}\ \eps(1,2,3,4),
\ee
with $4i\eps(1,2,3,4)=\ket{12}\vev{23}\ket{34}\vev{41}-\vev{12}\ket{23}\vev{34}\ket{41}$.
Through the order $\ap^8$ we find for any dimension $D$:
\begin{widetext}
\bea
%%  \lf.M(1,2,3,4,5)\ri|_{\ap^4}&=&\zeta(4)\ m(1,2,3,4,5)|_{\zeta(4)}+
%%  \lf(\fc{\pi^4}{120}\ s_{12}^5s_{34}+\fc{\pi^4}{36}\ s_{12}^3s_{34}^3+
%%   \fc{\pi^4}{120}\ s_{12}s_{34}^5\ri)\ \bar A_{YM}\ A_{YM}\nn\\
%%   &-&\zeta(2)^2\ \lf(s_{12}^3s_{34}+s_{12}s_{34}^3\ri)\ 
%%   \lf(\bar A_{F^4}\ A_{YM}+\bar A_{YM}\ A_{F^4}\ri)+\zeta(2)^2\ s_{12}s_{34}\ 
%% \bar A_{F^4}\ A_{F^4}\nn\\
%%  &+&{\rm permutations\ of\ } (23)=0\ ,\label{find54}\\[2mm]
\lf.M(12345)\ri|_{\ap^2}&=&0,\ \ \ \ \ \ \lf.M(12345)\ri|_{\ap^4}=0,\ \ \ \ \ \ 
\lf.M(12345)\ri|_{\ap^n}=\zeta(n)\ m(12345)|_{\zeta(n)}+(\overline{23})\ ,\ \ \ n=3,5,7\ ,\nnn
\lf.M(12345)\ri|_{\ap^6}&=&
\zeta(3)^2\ m(12345)|_{\zeta(3)^2}+
\zeta(3)^2\ \hat s_{12}\hat s_{34}\lf(\bar C_{1}\bar A_{YM} 
+\bar C_2\bar A_{F^4}\ri)|_{\zeta(3)}
\lf(C_{1} A_{YM}+C_2A_{F^4}\ri)|_{\zeta(3)}+(\overline{23}),\nnn
\lf.M(12345)\ri|_{\ap^8}&=&
\zeta(3)\zeta(5)\ m(12345)|_{\zeta(3)\zeta(5)}+
\zeta(3)\zeta(5)\  \hat s_{12} \hat s_{34}\;\lf[\lf(\bar C_{1}\bar A_{YM} 
+\bar C_2\bar A_{F^4}\ri)|_{\zeta(3)}
\lf(C_{1} A_{YM}+C_2A_{F^4}\ri)|_{\zeta(5)}\ri.\nn\\
&+&\lf.\lf(\bar C_{1}\bar A_{YM} 
+\bar C_2\bar A_{F^4}\ri)|_{\zeta(5)}\;\lf(C_{1} A_{YM}+C_2A_{F^4}\ri)|_{\zeta(3)}\ri]
+(\overline{23})\ .\label{find57}
\eea
\end{widetext}
Above we have introduced  ($\hat s_{ij}=(2\ap)^{-1}s_{ij}$):
\bea
m(12345)&=&\hat s_{12}\hat s_{34}\lf[\lf(\bar C_{1}\bar A_{YM} 
+\bar C_2\bar A_{F^4}\ri)A_{YM}\ri.\nn\\
&+&\lf.\bar A_{YM}\lf(C_{1} A_{YM}+C_2A_{F^4}\ri)\ri]\ .
\eea

\begin{figure}[H]
\centering
\includegraphics[width=0.45\textwidth]{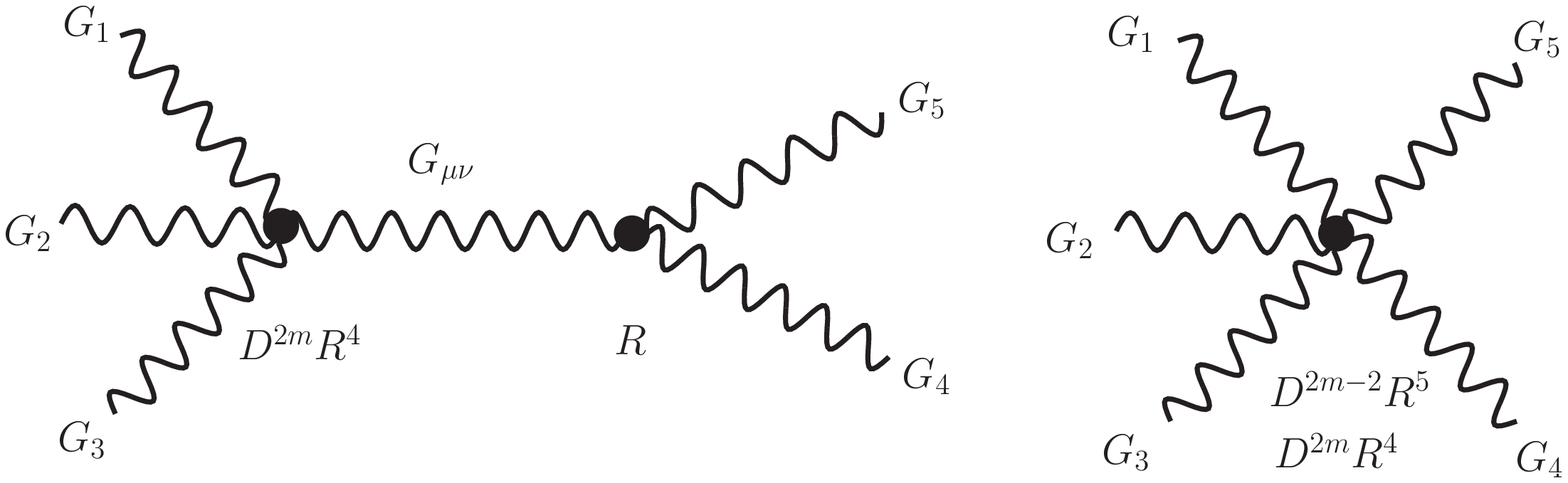}
\caption{\it Diagrams contributing at the order $\ap^{3+m}$ to the five--graviton amplitude.}
\label{reducible5}
\end{figure}

Let us now discuss the implications of the results \req{find57}.
In \cite{Russo:1998vt,Green:1999pv,Chandia:2003sh,Green:2005ba} 
the four--graviton amplitude has been analyzed resulting in 
a series of higher order terms $t_8t_8D^{2m}R^4$, which
enter Eq. \req{generic} with zeta functions 
shown in the first column of Table \ref{table}.
The (on-shell) four--graviton amplitude does not contain the momentum 
terms corresponding to $D^2R^4$ \cite{Metsaev:1987ju}.
The only possible Feynman diagrams contributing at the order $\ap^{3+m},m\!\geq\!0$ 
to the five--graviton amplitude are displayed in Fig. \ref{reducible5}.
For $m\!=\!0$ the above diagrams simply reproduce the $\ap^3$--order 
of the five--graviton amplitude involving the $R^4$ term.
\vskip-0.35cm
\vskip-0.4cm\noindent
Any $D^2R^4$ term  can always be rewritten as a sum of $R^5$ terms due 
to the relation $D^2R\!\simeq\!R^2$. 
Hence, for $m\!=\!1$ only the $5$--vertex (right diagram in Fig. \ref{reducible5}) 
stemming from an $R^5$ term may contribute at the order $\ap^4$
to the five--graviton amplitude. 
Since the latter vanishes at this order,  cf. \req{find57}, 
we conclude that an $R^5$ term does not exist.
At higher orders  $\ap^{3+m},m\!\geq\!2$ both $D^{2m-2}R^5$ and $D^{2m}R^4$ terms 
may contribute in  Fig.~\ref{reducible5}. 
The $\ap$--expansion of the five--graviton amplitude does not show
$\zeta(2)\zeta(3)$--terms at $\ap^5$, $\zeta(6)$--terms
at $\ap^6$, $\zeta(3)\zeta(4)$ nor $\zeta(2)\zeta(5)$--terms at $\ap^7$, and 
not any $\zeta(8)$, $\zeta(2)\zeta(3)^2$ and $\zeta(5,3)$--terms at $\ap^8$, cf.
Eqs.~\req{find57}. Since those terms cannot be generated 
by any reducible diagram with five external gravitons we conclude the absence of
contact interactions with coefficients $\ap^5\zeta(2)\zeta(3), \ap^6\zeta(6)$, 
$\ap^7\zeta(3)\zeta(4),\ap^7\zeta(2)\zeta(5)$, and $\ap^8\zeta(8),\ap^8\zeta(2)\zeta(3)^2,\ap^8\zeta(5,3)$, respectively, cf. second column of Table \ref{table}.
On the other hand, the non--vanishing terms  \req{find57}, which are 
proportional to $\ap^5\zeta(5),$ 
$\ap^6\zeta(3)^2$, $\ap^7\zeta(7)$, and $\ap^8\zeta(3)\zeta(5)$, respectively, account for the two diagrams in 
Fig. \ref{reducible5}. After subtracting the 
contribution of the left diagram the remaining terms give rise to combinations of
the interactions $D^4R^4, D^2R^5$ at $\ap^5$, combinations of 
$D^6R^4, D^4R^5$ at $\ap^6$, $D^ 8R^4,D^6R^5$ at $\ap^7$ and $D^{10}R^4, D^8R^5$ 
terms at $\ap^8$, respectively.
\begin{table}[h]
\begin{tabular}{|c||c|c|c|c|c|}\hline
&$N=4$ &$N=5$ &$N=6$ &$N=7$ &$N=8$ \\ \hline\hline
$\scriptstyle\ap^3\ \zeta(3)$ & $R^4$ & & & & \\ \hline
$\scriptstyle\ap^4\ \cancel{\zeta(4)}$ &  $\cancel{D^2R^4}$& $\cancel{R^5}$  & & & 
\\ \hline
$\scriptstyle\ap^5\ \zeta(5)$ & $D^4R^4$   & $D^2R^5$ & $R^6$ &  &\\
$\scriptstyle\ap^5\ \cancel{\zeta(2)\zeta(3)}$ &$\cancel{D^4R^4}$  & 
$\cancel{D^2R^5}$ & $\cancel{R^6}$ &&\\ \hline
$\scriptstyle\ap^6\ \zeta(3)^2$ &$D^6R^4$  &$D^4R^5$  &$D^2R^6$& $R^7\ ?$&\\ 
$\scriptstyle\ap^6\ \zeta(6)$ &$\cancel{D^6R^4}$  &$\cancel{D^4R^5}$ & 
$\cancel{D^2R^6}$& $R^7\ ?$&\\ \hline
$\scriptstyle\ap^7\ \zeta(7)$ &$D^8R^4$&$D^6R^5$  & $D^4R^6$ & $D^2R^7\ ?$ & $R^8\ ?$\\
$\scriptstyle\ap^7\ \zeta(3)\zeta(4)$ &$\cancel{D^8R^4}$&$\cancel{D^6R^5}$& 
$\cancel{D^4R^6}$ & $D^2R^7\ ?$ & $R^8\ ?$\\
$\scriptstyle\ap^7\ \zeta(2)\zeta(5)$ &$\cancel{D^8R^4}$ & $\cancel{D^6R^5}$ & 
$\cancel{D^4R^6}$ &$D^2R^7\ ?$ & $R^8\ ?$\\ \hline
$\scriptstyle\ap^8\ \zeta(3)\zeta(5)$ &$D^{10}R^4$&$D^8R^5$  & $D^6R^6$ & $D^4R^7\ ?$ & $D^2 R^8\ ?$\\
$\scriptstyle\ap^8\ \zeta(8)$ &$\cancel{D^{10}R^4}$&$\cancel{D^8R^5}$& 
$\cancel{D^6R^6}$ & $D^4R^7\ ?$ & $D^2R^8\ ?$\\
$\scriptstyle\ap^8\ \zeta(2)\zeta(3)^2$ &$\cancel{D^{10}R^4}$ & $\cancel{D^8R^5}$ & 
$\cancel{D^6R^6}$ &$D^4R^7\ ?$ & $D^2R^8\ ?$\\
$\scriptstyle\ap^8\ \zeta(5,3)$ &$\cancel{D^{10}R^4}$ & $\cancel{D^8R^5}$ & 
$\cancel{D^6R^6}$ &$D^4R^7\ ?$ & $D^2R^8\ ?$\\ \hline
\end{tabular}
\caption{\it Tree--level higher order gravitational couplings
and their corresponding zeta value coefficients probed by the $N$--graviton 
superstring amplitude.
Vanishing terms are crossed out. Those terms, which have not yet been 
probed by the relevant $N$--graviton amplitude,  are marked by a question mark.}
\label{table}\end{table}
\vskip-0.35cm\noindent

Next, we consider the scattering of six gravitons. The result for the six--gluon
subamplitude is given for any space--time dimension $D$ in \cite{DAN}, 
while in $D\ist4$ spinor notation in \cite{STi,STii,Potsdam}. Using 
these expressions  in  \req{kltii} yields the six--graviton 
amplitude.
The $\ap$--expansion of this amplitude gives vanishing results at the orders $\ap^2$
and~$\ap^4$:
\be\label{find6}
\hskip-0.05cm\lf.M(123456)\ri|_{\ap^2}=0,\ \lf.M(123456)\ri|_{\ap^4}=0.
\ee
The order $\ap^3$ is proportional to $\zeta(3)$ and describes
diagrams involving vertices from the $R^4$ coupling.
Moreover, through the order $\ap^8$ we find the following  properties: 
\be\hskip-0.25cm\ba{lll}
&\lf.M(123456)\ri|_{\zeta(2)\zeta(3)\ap^5}=0,
&\lf.M(123456)\ri|_{\zeta(6)\ap^6}=0,\\[1mm]
&\lf.M(123456)\ri|_{\zeta(2)\zeta(5)\ap^7}=0,
&\lf.M(123456)\ri|_{\zeta(3)\zeta(4)\ap^7}=0,\\[1mm]
&\lf.M(123456)\ri|_{\zeta(8)\ap^8}=0,
&\lf.M(123456)\ri|_{\zeta(2)\zeta(3)^2\ap^8}=0,\\[1mm]
&\lf.M(123456)\ri|_{\zeta(5,3)\ap^8}=0\ .&
\ea\label{find66}
\ee
Together with the previous results the findings \req{find66} restrict 
the contact interactions at $\ap^5$ to be of the form 
$\zeta(5)\{D^4R^4,D^2R^5,R^6\}$, but
forbids any $\zeta(2)\zeta(3)$ contact terms at this order. Similarly, contact interaction
at $\ap^6$ may assume the form $\zeta(3)^2\{D^6R^4,D^4R^5,D^2R^6\}$, but no  
interactions with $\zeta(6)$--factors are possible.
At this order also a reducible diagram describing the exchange of a graviton 
between two  four--vertices of $\zeta(3)R^4$ contributes. 
At the order $\ap^7$ only contact terms of the form 
$\zeta(7)\{D^8R^4,D^6R^5,D^4R^6\}$ may appear. However, no contact interactions
with $\zeta(3)\zeta(4)$ nor  $\zeta(2)\zeta(5)$--factors exist at this order in 
$\ap$. Eventually, at $\ap^8$ only contact terms of the form 
$\zeta(3)\zeta(5)\{D^{10}R^4,D^8R^5,D^6R^6\}$ may appear. However, no contact interactions 
with $\zeta(8), \zeta(2)\zeta(3)^2$ nor  $\zeta(5,3)$--factors exist at $\ap^8$. 
What remains to be checked is how for a given order in $\ap$ the set of contact 
interactions belonging to one row of Table \ref{table} can be expressed by a 
minimal basis of terms. The latter may allow to reduce the number of Riemann tensors
$R$ by converting them into derivatives $D^2$, cf. the comment \cite{commhigh}.

The results presented here and summarized in Table~\ref{table} hold for any type I or II superstring compactification in $D$ space--time dimensions with eight or more 
supercharges and suggest that higher order gravitational couplings 
\req{generic} obey some refined transcendentality properties: at each order in $\ap$
only  Riemann zeta functions of odd weight or products thereof appear.
While for the first column  this  is obvious to all orders in $\ap$ 
as a result of the relation \req{identzeta}, for the second and third column
we have checked this statement up to the order~$\ap^8$. Hence, 
for $n\!\leq\!6$ and up to order~$\ap^8$ the sum~\req{generic} 
runs only over basis elements comprised by MZVs of  odd weights \cite{Landon}.
The absence of the MZV $\zeta(5,3)$ at the order $\ap^8$
fits into this criterium, since it may be written as 
$\zeta(5,3)=-\fc{5}{2}\zeta(6,2)-\fc{21}{25}\zeta(2)^4+5\zeta(3)\zeta(5)$.

For $D\ist10$ type IIB superstring theory 
our findings together with the one--loop results \cite{Richards:2008jg} 
restrict the ring of possible modular forms describing 
the perturbative and non--perturbative completion of the higher order terms. 
%  Some terms of Table \ref{table} have been 
% investigated at one--loop by computing one--loop four and five graviton amplitudes
% \cite{Richards:2008jg} and at two--loops by computing two--loop four  graviton amplitudes
% \cite{DHoker:2005jc}.

The structure of the $\ap$--expansion takes over to amplitudes 
with some of the external gravitons replaced by some other member
of the supergravity multiplet.
E.g. in $D\ist4,\ \Nc=8$ the Fock space decomposition
$\lf|{\Nc=8\atop SUGRA}\ri\rng=\lf|{\Nc=4\atop SYM}\ri\rng
\otimes\lf|{\Nc=4\atop SYM}\ri\rng$ of the $256$ states of the 
$\Nc=8$ supergravity multiplet
selects the corresponding states in the gauge sectors, i.e. open string sectors.
In these sectors different amplitudes can be related  to all orders in $\ap$
by using supersymmetric Ward identities \cite{WARD}.
After applying the KLT relations the organization of the $\ap$--expansion 
stays the same as in the graviton case and the constraints for $D^{2m}R^n$ terms
take over to their supersymmetric variants. 

Our amplitude results, summarized in Table \ref{table}, have  impact on  
the recently discussed finiteness of $\Nc\ist8$ supergravity in $D\ist4$.
Counterterms invariant under $\Nc\ist8$ supergravity have an unique kinematic structure and
the tree--level string amplitudes provide  candidates for them, which are compatible 
with SUSY Ward identities and locality.
The absence or restriction on higher order gravitational terms at the order $\ap^l$
together with  their symmetries 
constrain the appearance of possible counter terms  available at $l$--loop, 
see \cite{Elvang:2010xn} for a  review and references therein.

{\bf Acknowledgments:}
I wish to thank Renata~Kallosh, and especially Lance Dixon for 
inspiring discussions and encouraging me to write up this article.
Furthermore, I thank 
Bernard de Wit,  Augusto Sagnotti, and especially Radu Roiban for discussions and 
Tobias Huber for his kind help to compute  one Euler integral and discussions.

%\vskip-0.75cm
\nocite{*}
\bibliography{r6}% Produces the bibliography via BibTeX.

\begin{thebibliography}{10}
\bibitem{commhigh}
Due to the relation $D^2R\!\simeq\! R^2$ some derivative terms may be converted
  to terms with fewer derivatives at the cost of higher orders in the Riemann
  tensor. Therefore, we stick to the prescription to write all terms with the
  highest possible number of Riemann tensors.
\bibitem{shuffle}
The set of integral linear combinations of MZVs is a ring, since the product of
  any two values can be expressed by a (positive) integer linear combination of
  the other MZVs \cite{Zagier}, e.g.:
  $\zeta(m)\zeta(n)=\zeta(m,n)+\zeta(n,m)+\zeta(m+n)$ (quasi--shuffle or
  stuffle relation).
\bibitem{Landon}
There are many relations over $\IQ$ among MZVs, e.g. $\zeta(4,1)\ist
  2\zeta(5)-\zeta(2)\zeta(3)$. For a given weight $w\in\IN$ the dimension $d_w$
  of the space spanned by MZVs of weight $w$ is given by $d_w=d_{w-2}+d_{w-3}$
  ($d_1=0,d_2=d_3=d_4=1,d_5=2,d_6=2,d_7=3,d_8=4,d_9=5,d_{10}=7,\ldots$)
  \cite{Zagier} and can be constructed by the following basis: for $w\ist 2$ by
  $\zeta(2)$, for $w\ist 3$ by $\zeta(3)$, for $w\ist 4$ by $\zeta(2)^2$, for
  $w\ist 5$ by $\zeta(5),\zeta(2)\zeta(3)$, for $w\ist 6$ by
  $\zeta(2)^3,\zeta(3)^2$, for $w\ist 7$ by
  $\zeta(7),\zeta(2)\zeta(5),\zeta(3)\zeta(2)^2$, for $w\ist8$ by $
  \zeta(2)^4,\zeta(2)\zeta(3)^2,\zeta(3)\zeta(5),\zeta(5,3)$, for $w=9$ by
  $\zeta(9),\zeta(7)\zeta(2),\zeta(5)\zeta(2)^2,\zeta(3)^3,\zeta(3)\zeta(2)^3$,
  for $w=10$ by
  $\zeta(7,3)$,$\zeta(5,3)\zeta(2),\zeta(7)\zeta(3),\zeta(5)^2,\zeta(5)\zeta(3%
)\zeta(2),\zeta(3)^2\zeta(2)^2,$ $\zeta(2)^5$, etc.
\end{thebibliography}
\bibliographystyle{h-physrev5}

\end{document}